\documentclass{w}
\twocolumn

\usepackage{gensymb}
\usepackage{amsmath}
\usepackage{amssymb}
\usepackage{colortbl}

\usepackage[OT2,OT1]{fontenc}
\newcommand\cyr
{
\renewcommand\rmdefault{wncyr}
\renewcommand\sfdefault{wncyss}
\renewcommand\encodingdefault{OT2}
\normalfont
\selectfont
}
\DeclareTextFontCommand{\textcyr}{\cyr}

\lat

\title{
Whispering gallery effect in relativistic optics}

\rtitle{
Whispering gallery effect in relativistic optics}

\sodtitle{
Whispering gallery effect in relativistic optics}

\author{Y.~Abe$^a$\thanks{e-mail: abe-y@ile.osaka-u.ac.jp},
 K.~F.~F.~Law$^a$,  Ph.~Korneev$^{b,c}$
\thanks{e-mail: korneev@theor.mephi.ru}, S.~Fujioka$^a $, S.~Kojima$^a $,  S.-H.~Lee$^a $, S.~Sakata$^a $,  K.~Matsuo$^a $,  A.~Oshima$^a $,  A.~Morace$^a $, Y.~Arikawa$^a $, A.~Yogo$^a $, M.~Nakai$^a $, T.~Norimatsu$^a $, E.~d'Humi\`eres$^{d}$, J.J.~Santos$^{d}$,  K.~Kondo$^e$,   A.~Sunahara$^f$,  S. Gus'kov$^{c,b}$ \thanks{e-mail: guskov@sci.lebedev.ru}, V. Tikhonchuk$^{d,g}$\\~\\} 

\address{
$^a $Institute of Laser Engineering, Osaka University, Japan\\
$^b$National Research Nuclear University MEPhI, Moscow, Russian Federation\\
$^c$P.~N.~Lebedev Physical Institute of RAS, Moscow, Russian Federation\\
$^d$University of Bordeaux, CNRS, CEA, CELIA, 33405 Talence, France\\
$^e$National Institute for Quantum and Radiological Science and Technology, Japan\\
$^f$Purdue University, USA\\
$^g$ELI-Beamlines, Institute of Physics Academy of Sciences of the Czech Republic, Doln\' i B\v re\v zany, Czech Republic}

\rauthor{Y.~Abe,  K.~F.~F.~Law,  Ph. Korneev et.al.}

\sodauthor{Y.~Abe,  K.~F.~F.~Law,  Ph. Korneev et.al.}

\abstract{A relativistic laser pulse, confined in a cylindrical target, performs multiple scattering along the target surface.  The confinement property of the target results in a very efficient interaction. This proccess, which is just yet another example of the ``whispering gallery'' effect, may pronounce itself in plenty of physical phenomena, including surface grazing electron acceleration and generation of relativistic magnetized plasma structures. }


\begin{document}

\maketitle

Development of laser technology opens bright horizons and at the same time poses new questions to physists. One of the most challenging, the high-intensity interaction regime, requires tight focusing of laser beams up to several wavelengths, and so a small-size target design and production becomes an important technological issue. An established branch of high-energy density laser-matter interaction physics combines geometrical and material properties of the targets used in order to absorb most of the laser energy and to redistribute it with the highest efficiency into directional electron or ion currents, shock waves, quasistatic fields, radiation, or other daughter effects. 
Several approaches have been already proved to be very efficient in terms of laser radiation absorption: nanoclusters \cite{Ditmire-pra96, Krainov-ufn00}, foams \cite{Limpouch-ppcf04, Okada-JJAP82, Chen-cisrep16}, structured targets \cite{Kahaly-prl08, Purvis-np13, Cristoforetti-ppcf14, Ivanov-lpl15}, or more tricky setups  \cite{Yang-apl17}. There, a target design defines predominant channels for efficient energy deposition.

In this Letter the ``whispering gallery'' effect, well known in acoustics \cite{Rayleigh-pms1910} and conventional optics, particularly in micro-resonators of semiconductor lasers \cite{Mccall-apl92}, for the first time reported as a phenomenon in relativistic optics; it was observed when an intense laser beam interacts with a dense surface plasma, generated inside a cylindrical-shaped target.  
By its geometrical properties, such a target possesses a possibility to capture the light, providing almost full laser pulse absorption. 
\begin{figure}{\begin{center}\includegraphics[width=0.5\textwidth]
{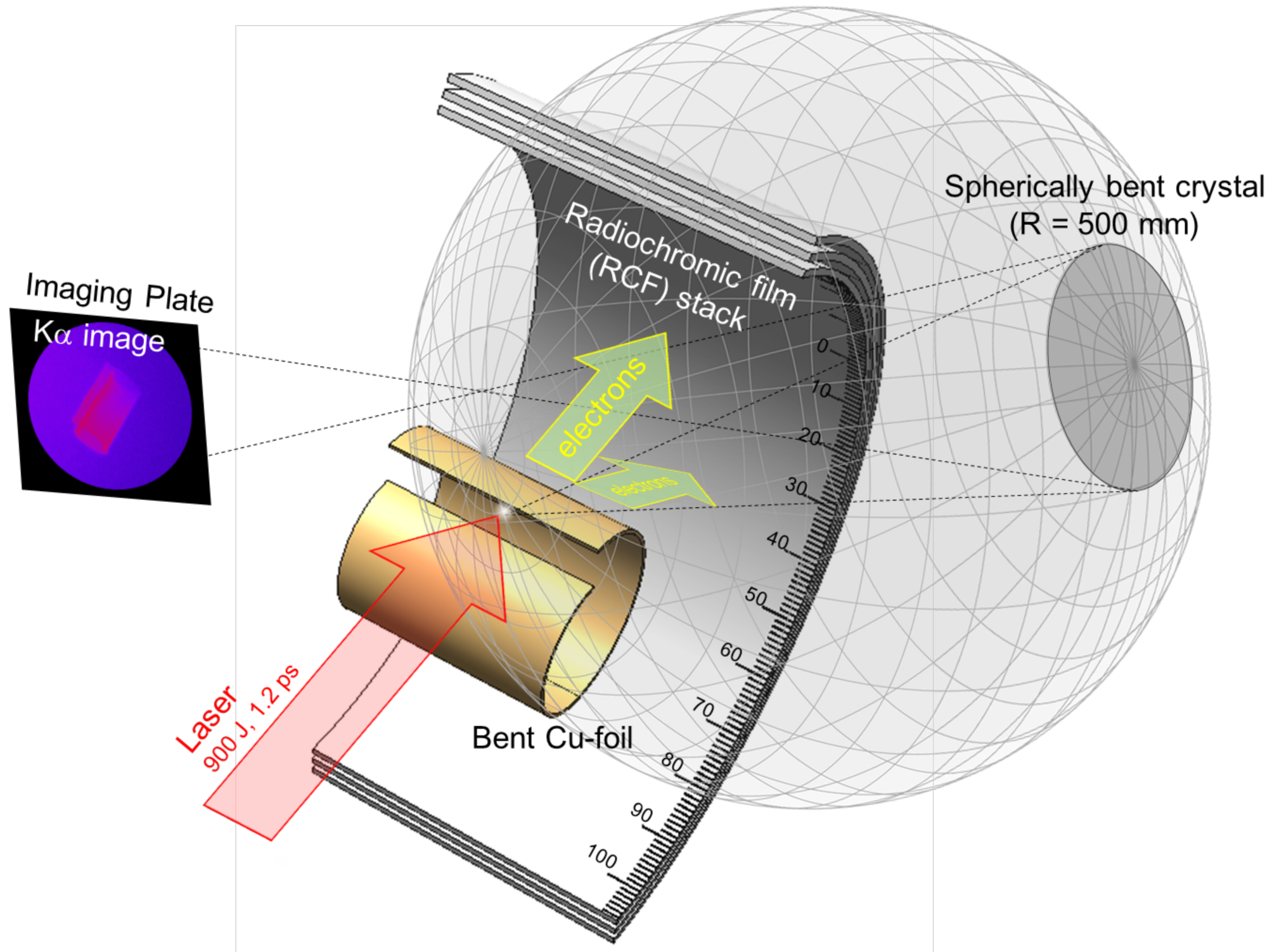}\end{center}}
\caption{A schematic sketch of the experimental setup. The main laser pulse, containing three equal beams of LFEX laser, enters the slit at grazing direction. The target is made from copper, to allow the Cu-Ka imaging of the hot electron with the use of a spherically bent crystal, shown behind the target. Also, a curved large-angle radiochromic film (RCF) stack, is placed around the target.}
\label{exp_setup}
\end{figure}

Those cylindrical-shaped, or ``snail'' targets were recently proposed for optical generation of super-intense quasistatic magnetic fields \cite{korneev-pre15}. It was shown in numerical studies, that the laser pulse, entering the curved target in the shape of a cylinder with an entrance slit, experiences multiple reflections, finally locked in the target internal volume. If the surface was an ideal material which permits total internal reflection, the Q-factor for corresponding whispering gallery modes would be very high. In our situation, the multiple reflection instead leads to a very high absorption and energy deposition, which in turn may results in many interesting accompanying effects. The one considered in the original work \cite{korneev-pre15} was the magnetic field generation, and deals with the very strong current generation along the target surface. An indispensable phenomena for such current generation was a very effective specular reflection. Though additional numerical studies  \cite{Korneev-jophcs2017} aimed to prove the robustness of the proposed generation scheme  were performed, the reduced target density, uncertainty in preplasma parameters, simplified material properties and other model assumptions could raise doubts concerning the demonstration of these results in experiments.


\begin{figure}{\begin{center}
\includegraphics[width=0.3\textwidth]
{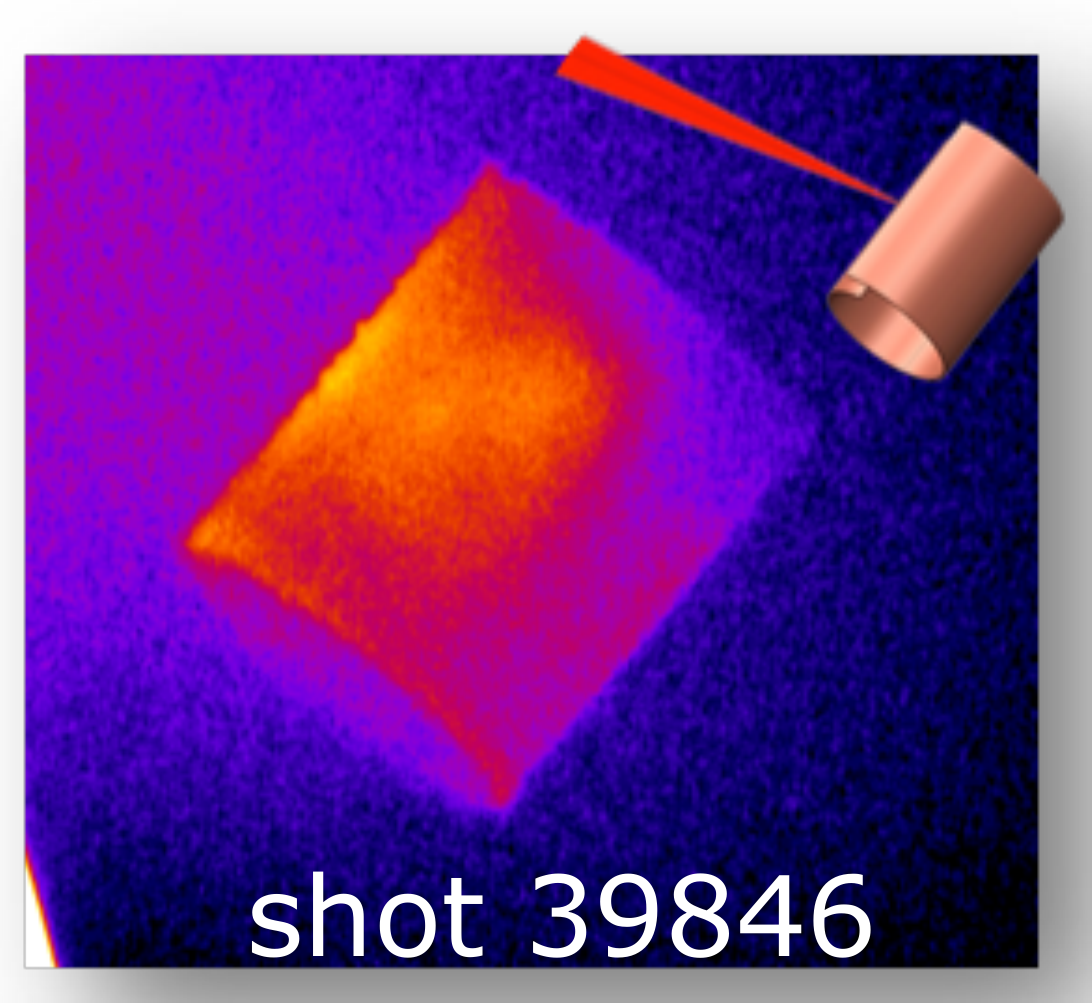}
\end{center}}
\caption{Image obtained by Cu-Ka emission from a snail target, irradiated by only the main laser pulse, which contained three LFEX beams, focused on the target surface through the entrance slit. The geometry of view is shown in the inset. The extended signal indicates a guiding of the laser beam inside the target. Assymetry between left and right edges of the target relates to laser alignment issues.}
\label{Cu-Ka}
\end{figure}

\begin{figure}{\begin{center}\includegraphics[width=0.5\textwidth]
{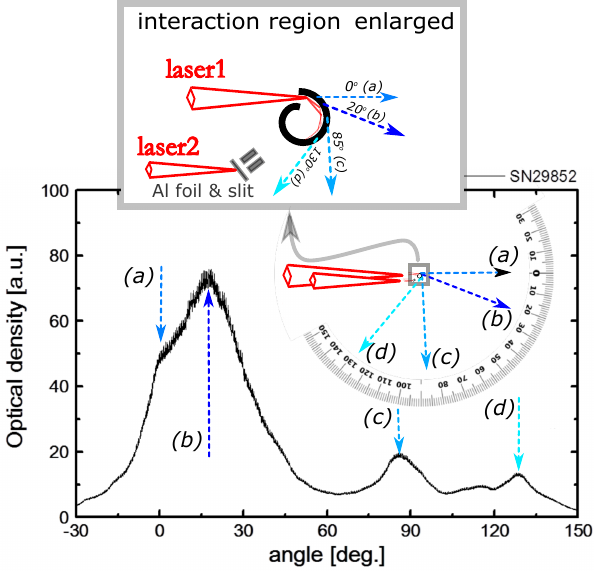}
\end{center}}
\caption{The optical density of electrons, collected by a large-angle, curved radiochromic film (RCF) stack. In this shot, two laser beams were used, and the two sources deposited the electrons to the RCFs, but the second (laser2) electron source was shielded by the slit, as shown in the inset. The first source, which was driven by  the laser1, produced consequent imprints of the laser pulse reflection inside the target. At each interaction zone this reflected beam generated a substantial relativistic electron flux, observed as a subsequent maxima in the figure at $\approx20\degree$, $\approx 85\degree$ and $\approx130\degree$. A strong maximum at $\approx0\degree$ corresponds to the direct electron acceleration. }
\label{fe}
\end{figure}

The relativistic ``whispering gallery'' effect was observed at the LFEX laser facility \cite{Miyanaga-jdp06} in the Institute of Laser Engineering (ILE) of the Osaka University.
The target was made from a copper foil of a 10 $\mu$m thickness, and had a curvature of 250 $\mu$m (diameter of 500 $\mu$m) at $0\degree$ decreasing down to 200 $\mu$m at $360\degree$, so having a 100 $\mu$m slit entrance for the laser beams, see a sketch of the experimental setup in Fig.~\ref{exp_setup}. The target length was 500 $\mu$m in the direction normal to the beam propagation.
Three beams of the LFEX laser were tightly focused on the inner surface of the curved target, entering the slit as shown in Fig.~\ref{exp_setup}.
The focal spot diameter of the combined laser beam was 60 $\mu$m, it contained 50\% of the total laser pulse energy, 710 $\pm$ 10 J. The laser pulse length was 1.3 $\pm$ 0.5 ps at full-width-at-half-maximum, so the laser intensity on the target was $\approx9.7 \times$ 10$^{18}$ W/cm$^2$, with the contrast between the main pulse intensity and the nano-second foot pulse better than 10$^{9}$ \cite{Fujioka-pp16}.

Two main diagnostics were used to demonstrate the effect of multiple successive surfing reflections of the laser beam inside the cylindrical target. The first one, based on the Cu-Ka emission due to the material heating by laser-accelerated electrons, was selected by a spherically-bent crystall with radius of 500 mm, and registered with image plates, as schematically shown in Fig.~\ref{exp_setup}. The image of Cu-Ka emission from the target, obtained in shot \# 39846 of LFEX laser, irradiated by three LFEX beams, is presented in Fig.~\ref{Cu-Ka}. The diagnostic image geometry is clarified by the inset, where the angle of view and the laser pulse direction are shown. As it follows from the diagnostic image, the interaction point is screched along the propagation direction, and follows the target surface. This may be explained by the laser continuous reflection inside the target. It may be noticed, that the left bottom part of the target is more bright in the registered spectrum. This is because the three LFEX beams were not focused absolutely symmetrically. Namely, the center of intensity appeared at left-bottom target side. 

Another diagnostic was performed by the radiochromic film (RCF) stack, positioned as shown in Fig. \ref{exp_setup}. The stack holder with a semi-cylindrical shape, covering the angular range from $-30\degree$ to $150\degree$ from the laser beam direction in the plane perpendicular to the axis of the cylinder, was mounted around the target to collect both electrons and ions from the target itself and the Al foil, used as a generator of probe protons under the fourth LFEX beam. This option was not used in the Cu-Ka image, shown in Fig. \ref{Cu-Ka}, but, because of the limited LFEX laser shots in the experiment, the option was ``on'' in the RCF imaging. However, due to a slit, mounted close to the Al foil, there is no direct image of this electron source on the RCF. Fig.~\ref{fe} shows the electron optical density, registered in the experiment, in the geometry, shown in the inset of Fig.~\ref{fe}. The optical density corresponds to the electron angular distribution, filtered from the ion signal. 
As it is evident in Fig.~\ref{fe}, there are several successive angular directions of the accelerated electrons. 

As it is shown in literature (see, e.g. \cite{Nakamura-prl04, Andreev-lpb14}), the observed electron signal may be explained by the interaction of the laser beam with a surface at grazing incidence. Thus we may deduce, that the set $\{\approx 20\degree;~\approx85\degree;~\approx130\degree \}$ relates to the laser beam internal reflections. Observed peak in electron distribution at $\approx 0\degree$ is formed by the fast electrons, generated by the non-reflected laser beam. 
Three internal reflections, as shown in Fig. \ref{fe}, evince that the laser pulse propagates along the target surface, like it does in a whispering gallery. In our setup, it propagates at least over a demisphere (due to geometry limitations, the RCF stack is limited to maximum angle of $150\degree$), which is almost 1 mm of the interaction length. Because of strong coupling, the intensity of the beam gradually decreases, though it is high enough to generate relativistic electrons at eachof the observed reflections.

The physics of interaction of an intense laser beam with confined curved targets, presented in this Letter, is very rich. Taking into account the grasing incidence, for the experimental conditions we may estimate the focal spot to be $\sim60\times120~\mu m$,  extended at least three times. With such nontrivial interaction conditions, we already observed the very efficient electron acceleration \cite{Korneev-arxiv17}, and strong current and magnetic field generation (in preparation). A rough estimate, with the laser energy $\sim700$ J and $\sim20-30\%$ energy conversion efficiency into hot electrons, results in $\sim100-200$ J in relativistic electrons. With the electron charge of $0.1$ mC and the pulse duration of $\sim1$ ps this correspond to $\sim100$ MA electric current, or the magnetic field in the range of tens of kT, in accordance with the predictions of numerical studies \cite{korneev-pre15}.

Concluding, we show the first experimental realization of the ``whisperring gallery'' effect in relativistic strong coupling regime of laser interaction with a cylindrically-bent foil. The multiple successive reflection of the laser beam inside the target volume leads inevitably to very efficient laser energy absorption -- there is no open channel left where the beam may escape. This in turn entails possibilities for a number of consequent effects. Some of these effects, such as the strong quasi-static magnetic field generation up to several kiloTesla and electron aceleration at curved surfaces with a very high energy cutoff were already studied in literature \cite{korneev-pre15, Korneev-arxiv17}, while the others are yet to be discovered.

The authors thank the technical support staff of ILE for assistance with the laser operation, target fabrication, plasma diagnostics, and computer simulations. 
This work was supported by the Collaboration Research Program of ILE (2016Korneev and 2017Korneev); by the Japanese Ministry of Education, Science, Sports, and Culture (MEXT) and Japan Society for the Promotion of Science (JSPS) through Grants-in-Aid for Scientific Research (A) (No. JP16H02245); by Challenging Exploratory Research (No. JP16K13918); by the Bilateral Program for Supporting International Joint Research between Japan and Russia by JSPS.
This work was supported by the Russian Foundation for Basic Research (\# 16-52-50019\,\hskip-9pt{\cyr{YaF}}), Excellence programm NRNU MEPhI. 


\end{document}